\documentclass[aps,prb,showpacs]{revtex4}
\usepackage{graphicx,amsmath,amssymb,color}
\usepackage{float}
\usepackage{epstopdf}
\begin{document}
\title{Effect of Energy Band Gap in Graphene  on Negative Refraction through the Veselago Lens
and Electron Conductance}
\author {Dipendra Dahal$^{1}$  and  Godfrey Gumbs$^{1,2}$ }
\affiliation{$^{1}$Department of Physics and Astronomy, Hunter College of the
City University of New York, 695 Park Avenue, New York, NY 10065, USA\\
$^{2}$Donostia International Physics Center (DIPC),
P de Manuel Lardizabal, 4, 20018 San Sebastian, Basque Country, Spain}

\begin{abstract}

A remarkable property of intrinsic graphene is that upon
doping, electrons and holes travel through the monolayer  thick material with
constant velocity which does not depend on energy up to about $0.3$ eV  (Dirac fermions),
 as though the electrons and  holes are massless particles and antiparticles which move
at the Fermi velocity $v_F$.  Consequently, there is Klein tunneling at a $p-n$  junction,
in which there is no backscattering at normal incidence of massless Dirac fermions.
However, this process  yielding perfect transmission at normal incidence is expected to
be affected  when the group velocity of the charge carriers
is energy dependent and  there is non-zero effective mass   for the target particle. We
investigate how away from normal incidence  the combined effect of incident
electron energy $\epsilon$ and  band gap  parameter $\Delta$  can determine
whether a $p-n$ junction would  allow focusing of an electron beam by behaving
like a Veselago lens with  negative  refractive index. We demonstrate that
there is a  specific region in $\epsilon-\Delta$  space where the index of
refraction is negative, i.e., where monolayer graphene behaves as a metamaterial. 
Outside this region, the   refractive index may be positive or
there may be no refraction at all.  We compute the ballistic conductance across a $p-n$
junction as a function of  $\Delta$ and $\epsilon$ and compare  our results with
those for a single electrostatic potential barrier and multiple barriers.

\end{abstract}

 \pacs{73.20.-r, \ 73.20.Mf, \ 78.20.Bh, \ 78.67.Bf}

\maketitle

\section{  Introduction}
\label{sec1}

Graphene is a virtually two-dimensional (2D) sheet of carbon atoms which is
 nearly transparent and a considerably strong material for its light weight,
 high thermal and electrical conductivity. It is an allotrope of carbon atoms
 with 2D properties. The atoms are packed densely in a regular $sp^2$ bonded
 atomic scale chicken wire hexagonal pattern. Creating high quality graphene
 is a complex process which prevented it from being easily available \cite{exp1,exp2,exp3}.
However, recent work found that by analyzing graphene's interfacial adhesive
energy, it is possible to separate graphene from the metallic background on which
it is grown. With the experimental realization of graphene, the focus  now is to
obtain  a thorough understanding of  its  electronic and photonic properties. In
this regard, one of the most intriguing challenges of graphene is
a comprehensive  understanding of  the transmission \cite{trans1, trans2, trans3, trans4, trans5, trans6, trans7}
of charged  particles across a  potential barrier or a potential step.

\medskip
\par
In the case for transmission of an  electron through a higher barrier than its incoming energy,    the electron turns into a hole within the barrier region,
 with its momentum directed in the  reversed direction, resulting in a negative refractive
  index \cite{Been, Negrf1, Negrf2, Negrf3, Negrf4, Negrf5, Negrf6, Negrf7, Negrf8} causing the electron beam to focus at a point, thereby producing a Veselago
   lens.    There is always some probability for the electron to tunnel onto the other
   side of the barrier, which we have calculated in this paper.    Our transmission
   probability is then employed in our calculation of the  conductance \cite{cond1,cond2, cond3}.
The behavior    of the transmission coefficient as well as   the conductance
 changes dramatically when an energy band gap is introduced in  graphene which may be
 achieved either by placing the 2D monolayer on a substrate or exposing it to
  circularly polarized light \cite{band1, band2}.

\medskip
\par
The rest of our paper is organized as follows.
  In Sec. \ref{sec2},     the $\sigma_z$ model Hamiltonian is used to solve the
  Hamiltonian equation for    the wavefunction in the various regions with a prescribed constant potential.
   Using the continuity condition for the wavefunction at an   interface between regions, the
   transmission probability is calculated. This is carried out
for a  potential step,   as well as for one and two electrostatic potential barriers. In
Sec.\ \ref{sec3}, we present numerical results for the transmission coefficient, and the
conductance using the analytic results in Sec.\ \ref{sec2}. We conclude our paper with some
relevant remarks in Sec.\ \ref{sec4}.

\medskip
\par

\section{Theoretical Model and Formulation}
\label{sec2}

\subsection{Single  Potential Step:  \ n-p junction}

\begin{figure}[h]
\centering
\includegraphics[width=2.87in,height=3.21in,keepaspectratio]{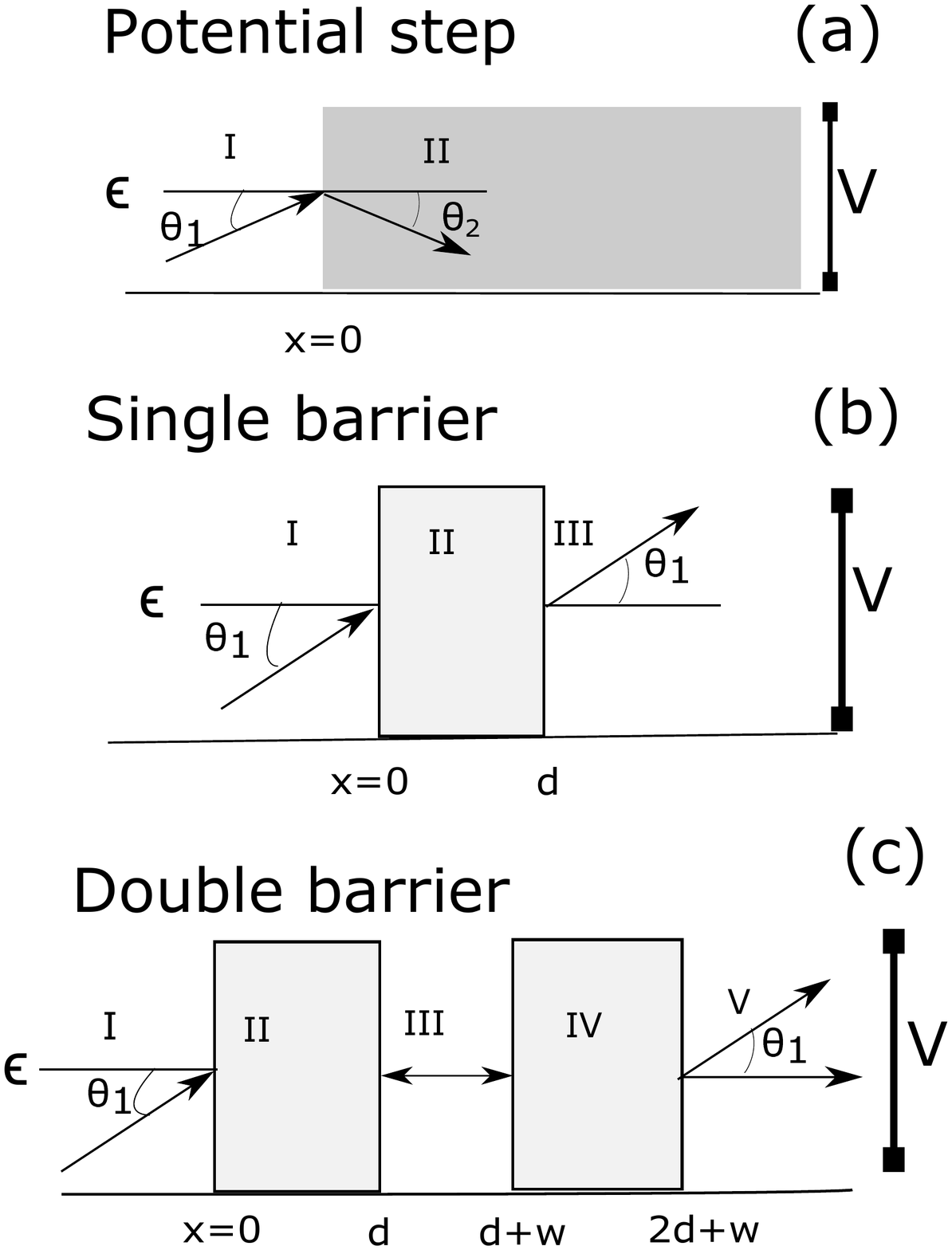}
\caption{(Color online) Schematic representation of (a) a step potential, $\epsilon<V$,
(b) a single potential  barrier, and (c) a pair of potential barriers.  }
\label{FIG:a}
\end{figure}

First, we consider  Dirac fermions with energy $\epsilon$ incident from the left-hand side of
an  electrostatic potential barrier at an  angle $\theta_1$ with respect to $x$-axis.
This structure may be produced by applying a local top gate voltage to  graphene.
The model consists of regions ``I" and ``II", with  ``II"  denoting the step
region and ``I"  the entry region into  the potential step. The interface
between the two regions  is located at $x=0$ as shown schematically in Fig.\
\ref{FIG:a}(a).

\medskip
\par

At low energy, the electronic properties of monolayer gapped graphene,  although
the microscopic  Hamiltonian of carbon atoms  is non-relativistic,
  are governed by a Dirac Hamiltonian given by
	
\begin{equation}
	{\cal H}=\begin{pmatrix}\Delta +V (x) &  v_F(\hat p_x-i\hat p_y) \cr
	 v_F(\hat p_x+i \hat p_y) & -\Delta+V (x)\end{pmatrix} \ ,
\end{equation}
where $\hat p_\ell=-i\hbar\partial/\partial x_\ell$ with $x_\ell=x,y$ and
$v_F$  is the Fermi velocity. The energy  eigenvalues  are given by
$\epsilon=V_\alpha \pm \sqrt{(\hbar v_F q_\alpha)^2+\Delta^2}$ in region
 $\alpha = $ 1, 2 with constant potential $V_\alpha$ and
 $q_\alpha=\sqrt{q_{\alpha,x}^2+q_{\alpha,y}^2}$ is the magnitude of the wave vector
 in that region.    Also,  $\Delta$ is a parameter  describing the energy band gap.
Let $\theta_1$ be the angle of incidence of the incoming electron with wave vector $q_1$
in region I and $\theta_2$ is the angle of refraction in region II.
The relationship between $\theta_1$ and $\theta_2$ is  given by

\begin{equation}
\frac{\tan\theta_1}{\tan\theta_2}=s_2 \frac{|q_{2,x}|}{|q_{1,x|}}= s_2
\frac{\sqrt{\frac{(\epsilon-V_2 )^2-\Delta^2}{\hbar^2 v_F^2}-q_{2,y}^2}}
{\sqrt{\frac{\epsilon^2-\Delta^2}{\hbar^2v_F^2}-q_{1,y}^2}}\ ,
\end{equation}

\begin{equation}
\frac{\sin\theta_1}{\sin\theta_2}=\frac{q_{2}}{q_{1}}
= s_2 \sqrt{\frac{(\epsilon-V_2 )^2-\Delta^2}{\epsilon^2-\Delta^2}  }
\label{angle1} \ ,
\end{equation}
where $s_\alpha={\mbox sgn}(\epsilon-V_\alpha)$, i.e. when $\epsilon<V_\alpha$, we
are able to get negative refraction \cite{Negrf5, Negrf6, Negrf7, Negrf8}.

Making use of  the Heisenberg equation of motion, the velocity operator can be evaluated as

\begin{equation}
\hat{v}=\dot{\bf  r}=\frac{[{\bf r},{\cal H}] }{i}={\hat{\bf \sigma}} \ .
\end{equation}
The probability density for the state $|\psi>$ is $|\psi|^2$ and the average probability current is
denoted by $\bf j(r,t)$. The conservation of probability is given by

\begin{equation}
\nabla\cdot {\bf j}+\frac{d|\psi|^2}{dt}=0 \ .
\end{equation}
As the wave function is time-independent, we have $\nabla \cdot \bf j=0$
and $\bf j=\Psi^\dag\hat{\sigma}\Psi$.
Also, the wavefunction in region   $\alpha$ is represented by

\begin{equation}
\Psi_\alpha(x,y)=\left\{ a_\alpha e^{iq_{\alpha x}x}\begin{pmatrix}1\\g_\alpha e^{i\theta_\alpha}\end{pmatrix}
+r_\alpha e^{-iq_{\alpha x}x}\begin{pmatrix}1\\g_\alpha e^{i(\pi-\theta_\alpha)}\end{pmatrix}\right\}
e^{iq_{\alpha y}y}   \ ,
\end{equation}
where

\begin{equation}
g_\alpha =  s_\alpha \left\{ \frac{|\epsilon-V_\alpha|-\Delta}{\sqrt{(\epsilon-V_\alpha)^2
-\Delta^2}}\right\} \ .
\end{equation}
Additionally, when  $\epsilon<V_2$, inside the barrier region, the carrier electron is now in the valence band,
transformed into a hole  and  for  $\epsilon>V_2$, the particle is in the conduction
band and maintains itself as an electron. For a step,  we set     $V_1$=$0$ and $V_2$=$V$.
Using the continuity of the wave function at the interface and taking $a_1$=$1$,  $r_2$=$0$ and
$a_2$=$t$, we have

\begin{eqnarray}
1+r_1&=& t \ ,
\nonumber\\
g_1e^{i\theta_1}+r_1 g_1e^{i\pi-i\theta_1} &=& tg_2e^{i\theta_2} \ .
\end{eqnarray}
Solving these simultaneous equations for $t$. we obtain

\begin{equation}
|t|^2=\frac{4 g_1^2 \cos^2\theta_1}{g_1^2+g_2^2+2 g_1g_2\cos(\theta_1+\theta_2)} \ .
\end{equation}
In a state of stable equilibrium,    the continuity equation yields current conservation
which  implies that

\begin{equation}
j_x(\mbox{incident})+j_x(\mbox{reflected})=j_x(\mbox{transmitted})
\end{equation}
from which we obtain

\begin{equation}
g_1\cos\theta_1-r_1^2g_1\cos\theta_1=t^2g_2\cos\theta_2 \ ,
\end{equation}
or

\begin{equation}
1=r_1^2+t^2 \frac{g_2\cos\theta_2}{g_1\cos\theta_1} \ .
\end{equation}
From  the conservation of probability, we have  $R+T=1$,
where $T$ is the  transmission probability given by

\begin{equation}
T=|t|^2\frac{g_2\cos\theta_2}{g_1\cos\theta_1}
\end{equation}
and $R$ is the reflection coefficient.
Therefore,   for gapped graphene,  this gives us

\begin{equation}
T=\frac{4g_1g_2\cos\theta_1\cos\theta_2}{g_1^2+g_2^2+2g_1g_2\cos(\theta_1+\theta_2)} \ .
\end{equation}
When $\epsilon>V_2$ and the band gap is zero,  this  reduces to

\begin{equation}
T=\frac{2\cos\theta_1\cos\theta_2}{1+\cos(\theta_1+\theta_2)} \ .
\end{equation}
However,  when $\epsilon<V_2$  for gapless graphene, the angle
$\theta_2$ is replaced by   $\theta_2+\pi$.and  we have instead

\begin{equation}
T=-\frac{2\cos\theta_1\cos\theta_2}{1-\cos(\theta_1+\theta_2)}
\end{equation}
which agrees with the result in Ref.\ [\onlinecite{Negrf8}].

\medskip
\par

\subsection{Single Potential Barrier: \ n-p-n junction}

We now turn our attention to a  model consisting of regions ``I", ``II" and ``III",
as shown in Fig.\ \ref{FIG:a}(b), with  ``II"  denoting the barrier
region and ``I" and ``III" the regions of entry and exit from the barrier,
 respectively. The boundaries of the barrier are located at $x=0$ and $x=d$.
Let $\theta_1$ be the angle of incidence for the incoming electron with wave vector $q_1$
in region I. The corresponding wave functions for the electron  in the three regions are

\begin{eqnarray}
\Psi_I(x,y) &=&\left\{ A\begin{pmatrix}1\\
g_1e^{i\theta_1}\end{pmatrix}e^{iq_{1,x}x}+B\begin{pmatrix}1\\
g_1e^{i\pi-i\theta_1}\end{pmatrix}e^{-iq_{1,x}x}\right\} e^{iq_{1,y}y} \ ,
\nonumber\\
\Psi_{II}(x,y) &=&\left\{ C\begin{pmatrix}1\\
g_2e^{i\theta_2}\end{pmatrix}e^{iq_{2,x}x}+D\begin{pmatrix}1\\ g_2e^{i\pi-i\theta_2}
\end{pmatrix}e^{-iq_{2,x}x}\right\} e^{iq_{2,y}y} \ ,
\nonumber\\
\Psi_{III}(x,y) &=&\left\{ E\begin{pmatrix}1\\
g_1e^{i\theta_1}\end{pmatrix}e^{iq_{1,x}x}+F\begin{pmatrix}1\\ g_1e^{i\pi-i\theta_1}
\end{pmatrix}e^{-iq_{1,x}x}\right\} e^{iq_{1,y}y} \ .
\end{eqnarray}
In our notation, $A, B, C,D,E$ and $F$ are constants,
$q_2^2=\frac{(\epsilon-V_2)^2-\Delta^2}{ (\hbar^2 v_F^2)}$ is the
square of the wave vector within the potential barrier
$V_2=V$ and, $\epsilon$ is the energy of the incident electron.  Also,
$\theta_1$ is the   angle which the incoming electron  makes with the $x$-axis
 and $\theta_2$ is the refracted angle inside the barrier. We have

\begin{equation}
\theta_2=\tan^{-1}\left(\frac{q_2 \sin\theta_2}{\sqrt{(\epsilon-V_2)^2-\Delta^2-(q_2 \sin\theta_2)^2}}
\right) \
\label{angle}
\end{equation}

\medskip
\par

We now consider the case when there is a potential barrier ($n-p-n$ junction)
for gapped graphene and the continuity of the wave function gives

\begin{equation}
{\cal M}_1\begin{pmatrix}
A\cr
B
\end{pmatrix}={\cal N}_1
\begin{pmatrix}C\\D\end{pmatrix} \ ,
\end{equation}
where,

\begin{eqnarray}
{\cal M}_1&=& \begin{pmatrix}1&&1\\ g_1 e^{i\theta_1}&&g_1 e^{i\pi-i\theta_1}\end{pmatrix} \ ,
\nonumber\\
{\cal N}_1&=&\begin{pmatrix}1&&1\\ g_2 e^{i\theta_2}&&g_2 e^{i\pi-i\theta_2}\end{pmatrix} \ .
\end{eqnarray}
Transforming to a primed coordinate system  by translating the origin  to a
new one at $x=d$, i.e. $x'=x-d$, we have,  for region ``II"

\begin{eqnarray}
\Psi_{II}(x,y) &=& \left\{C \begin{pmatrix}1 \\ g_2 e^{i\theta_2}\end{pmatrix}e^{i  q_{2,x} x'}e^{iq_{2,x}d}+D\begin{pmatrix}1 \\ g_2 e^{i\pi-i\theta_2}\end{pmatrix}e^{-i q_{2,x} x^\prime}e^{iq_{2,x}d}
\right\}e^{iq_{2,y}y} ,
\nonumber\\
&=& \left\{C' \begin{pmatrix}1 \\ g_2 e^{i\theta_2}\end{pmatrix}e^{i  q_{2,x} x'}+D'\begin{pmatrix}1 \\ g_2 e^{i\pi-i\theta_2}\end{pmatrix}e^{-i q_{2,x} x^\prime}\right\}e^{iq_{2,y}y} \ ,
\end{eqnarray}
where

\begin{eqnarray}
\begin{pmatrix}C'\\D'\end{pmatrix}&=& \begin{pmatrix}Ce^{iq_{2,x} d}\\De^{-iq_{2,x}d}\end{pmatrix}
=\begin{pmatrix}e^{iq_{2,x}d}&&0\\0&&e^{-iq_{2,x} d}\end{pmatrix}
\begin{pmatrix}C\\D\end{pmatrix} \ .
\end{eqnarray}
Similarly, in  region III,

\begin{eqnarray}
\Psi_{III}(x,y)&=&\left\{E'\begin{pmatrix}1 \\ g_1 e^{i\theta_1}
\end{pmatrix}e^{iq_{1,x}x'}+F'\begin{pmatrix}1 \\ g_1 e^{i\pi-i\theta_1}\end{pmatrix}e^{-iq_{1,x}x'}\right\}e^{iq_{1,y}y} \  ,
\end{eqnarray}

\begin{eqnarray}
\begin{pmatrix}E'\\ F'\end{pmatrix}&=& \begin{pmatrix}Ee^{i q_{1,x} d}\\Fe^{-iq_{1,x}d}\end{pmatrix}=
\begin{pmatrix}e^{iq_{1,x}d}&&0\\0&&e^{-iq_{1,x}d}\end{pmatrix}
\begin{pmatrix}E\\F\end{pmatrix} \ ,
\end{eqnarray}
where

\begin{eqnarray}
{\cal T}_{10}&=&\begin{pmatrix} e^{iq_{1,x}d}&&0 \\
0&&e^{-iq_{1,x}d} \end{pmatrix} \ ,
\nonumber\\
{\cal T}_{1V} &=&\begin{pmatrix} e^{iq_{2,x}d}&&0 \\
0&&e^{-iq_{2,x}d} \end{pmatrix} \ .
\end{eqnarray}
In the primed coordinate frame of reference, we have

\begin{equation}
\begin{pmatrix}1&&1\\ g_2 e^{i\theta_2}&&g_2 e^{i\pi-i\theta_2}\end{pmatrix}\begin{pmatrix}
C'\cr
D'
\end{pmatrix} = \begin{pmatrix}1&&1\\ g_1
e^{i\theta_1}&&g_1 e^{i\pi-i\theta_1}\end{pmatrix}\begin{pmatrix}E'\\F' \ ,
\end{pmatrix}
\end{equation}
\begin{equation}
\begin{pmatrix}C'\\D'\end{pmatrix}  =  {\cal N}_1^{-1}.{\cal M}_1.\begin{pmatrix}E'\\F'
\end{pmatrix} \ ,
\end{equation}
where, in our notation,

\begin{eqnarray}
\nonumber\\
\begin{pmatrix}C\\D\end{pmatrix} &=&  {\cal T}_{1V}^{-1}.{\cal N}_1^{-1}.{\cal M}_1.{\cal T}_{10}.\begin{pmatrix}E\\F\end{pmatrix} \ ,
\nonumber\\
\begin{pmatrix}
A\cr
B\end{pmatrix}
&=&{\cal M}_1^{-1}.{\cal N}_1.{\cal T}_{1V}^{-1}.{\cal N}_1^{-1}.{\cal M}_1.{\cal T}_{10} \ .
  \begin{pmatrix}E\cr
F\end{pmatrix} \ .
\end{eqnarray}
Setting  $A=1$, $E=t$ and $F=0$, we obtain the  transmission probability
$T=|t|^2$ given by

\begin{equation}
T=\frac{4 g_1^2g_2^2 \cos^2\theta_1 \cos^2\theta_2}{4g_1^2g_2^2 cos^2\theta_1 cos^2\theta_2\cos^2(dq_2)+\sin^2(dq_2)(g_1^2+g_2^2-2g_1g_2 \sin\theta_1 sin\theta_2)^2} \ ,
\label{general}
\end{equation}
This general result agrees with that in Ref. [\onlinecite{trans1}]
 for the special case of gapless graphene and we shall employ Eq.\ (\ref{general})
 in our numerical calculations presented below.

\subsection{Two Potential Barriers }

We now  extend our formulation to  calculations for two square potential  barriers of
equal height $V$, illustrated in Fig.\ \ref{FIG:a}(c)
The first barrier is located between $x=0$ and $x=d$ (region II) while the second
barrier lies between $x=d+w$   and $x=2d+w$ (region IV). The region between the
barriers is denoted by region III, and to the far left and far right we
have regions I and V, respectively. A straightforward calculation leads to
the wave function for  region III as

\begin{eqnarray}
\Psi_{III}(x,y)&=&\left\{ E \begin{pmatrix}1 \\ g_1 e^{i\theta_1}\end{pmatrix}e^{iq_{1,x}x}+F\begin{pmatrix}1 \\ g_1 e^{i\pi-i\theta_1}\end{pmatrix}e^{-iq_{1,x}x}\right\}e^{iq_{1,y}y}  \ .
\end{eqnarray}
Making use of,

\begin{eqnarray}
\begin{pmatrix}E''\\ F''\end{pmatrix}&=& \begin{pmatrix}Ee^{i q_{1,x} (d+w)}\\Fe^{-iq_{1,x}(d+w)}\end{pmatrix}=
\begin{pmatrix}e^{iq_{1,x}(d+w)}&&0\\0&&e^{-iq_{1,x}(d+w)}\end{pmatrix}
\begin{pmatrix}E\\F\end{pmatrix} \ ,
\end{eqnarray}
we  obtain,

\begin{eqnarray}
\Psi_{III}(x,y)&=&\left\{E''\begin{pmatrix}1 \\ g_1 e^{i\theta_1}
\end{pmatrix}e^{iq_{1,x}x''}+F''\begin{pmatrix}1 \\ g_1 e^{i\pi-i\theta_1}\end{pmatrix}
e^{-iq_{1,x}x''}\right\}e^{iq_{1,y}y} \  .
\end{eqnarray}
In a similar fashion, the wave function in the second barrier region (region IV)
may be written as

\begin{eqnarray}
\Psi_{IV}(x,y)&=& \left\{  G''\begin{pmatrix}1 \\g_2 e^{i\theta_2}
\end{pmatrix}e^{iq_{2,x}x''}+H''\begin{pmatrix}1 \\g_2 e^{i\pi-i\theta_2}\end{pmatrix}e^{-iq_{2,x}x''}\right\}e^{-iq_{2,y}y} \ ,
\end{eqnarray}
and

\begin{equation}
  \begin{pmatrix}
G^{\prime\prime}\cr
H^{\prime\prime}\end{pmatrix}
= \begin{pmatrix}
e^{iq_{2,x}(d+w)}&0\cr
0& e^{-iq_{2,x}(d+w)}\end{pmatrix}
   \begin{pmatrix}G\cr
H\end{pmatrix} \ ,
\label{t20}
\end{equation}
where $G, H$ are constants and $x^{\prime\prime}=x-(d+w)$. The wave function in the second barrier is
 also continuous at $x^{\prime\prime}  =0$ i.e., at $x=d+w$ we must have:

 \begin{equation}
\begin{pmatrix}E^{\prime\prime}\cr
F^{\prime\prime}\end{pmatrix} ={\cal M}_1^{-1}. {\cal N}_1. \begin{pmatrix}G''\cr
H^{\prime\prime}\end{pmatrix}
\end{equation}
which leads to

\begin{equation}
  \begin{pmatrix}E\cr
F\end{pmatrix} ={\cal T}_{20}^{-1}.{\cal M}_1^{-1}. {\cal N}_1. {\cal T}_{2V}.
\begin{pmatrix}G\cr H\end{pmatrix} \ ,
\end{equation}
with

\begin{eqnarray}
{\cal T}_{20}&=&\begin{pmatrix} e^{iq_{1,x}(d+w)}&&0 \\
0&&e^{-iq_{1,x}(d+w)} \end{pmatrix} \ ,
\nonumber\\
{\cal T}_{2V}&=&\begin{pmatrix} e^{iq_{2,x}(d+w)}&&0 \\
0&&e^{-iq_{2,x}(d+w)} \end{pmatrix} \ .
\end{eqnarray}
Proceeding along lines like   we employed  above, we obtain the
components of the  wave function in region V as

\begin{eqnarray}
\Psi_{V}(x,y)&=&\left\{ J\begin{pmatrix}1 \\ g_1 e^{i\theta_1}\end{pmatrix}
e^{iq_{1,x}x}+K\begin{pmatrix}1 \\ g_1 e^{i\pi-i\theta_1}\end{pmatrix}
e^{-iq_{1,x}x}\right\}e^{iq_{1,y}y} \ ,
\end{eqnarray}
where $J$ and $K$  are constants. This  may be rewritten as

\begin{eqnarray}
\Psi_{V}(x,y) &=&\left\{ J^{iv}\begin{pmatrix}1 \\ g_1 e^{i\theta_1}\end{pmatrix}e^{iq_{1,x}x^{iv}}+K^{iv}\begin{pmatrix}1 \\ g_1 e^{i\pi-i\theta_1}\end{pmatrix}e^{-iq_{1,x}x^{iv}}\right\}e^{iq_{1,y}y} \ ,
\end{eqnarray}
where $x^{iv}=x-(2d+w)$. Furthermore,  in   region IV, the
wave function may be expressed in the following form:

\begin{eqnarray}
\Psi_{IV}(x,y)&=&\left\{ G^{iv}\begin{pmatrix}1\\e^{i\theta_2}\end{pmatrix}e^{iq_{2,x}x^{iv}}
+H^{iv}\begin{pmatrix}1\\e^{i\pi-i\theta_2}\end{pmatrix}e^{-iq_{2,x}x^{iv}}\right\} e^{iq_{2,y}y}
\end{eqnarray}

with

\begin{eqnarray}
\begin{pmatrix}J^{iv}\cr K^{iv}\end{pmatrix}&=& \begin{pmatrix}e^{iq_{1,x}(2d+w)}& 0\cr
0&e^{-iq_{1,x}(2d+w)}\end{pmatrix}\begin{pmatrix}J\cr
K\end{pmatrix}\equiv  {\cal T}_{30}\begin{pmatrix}J\cr
K\end{pmatrix}
\nonumber\\
\begin{pmatrix}G^{iv}\cr
H^{iv}\end{pmatrix}&=& \begin{pmatrix}e^{iq_{2,x}(2d+w)} &0\cr
0 &e^{-iq_{2,x}(2d+w)}\end{pmatrix}\begin{pmatrix}G\cr
H\end{pmatrix} \equiv   {\cal T}_{3V}\begin{pmatrix}G\cr
H\end{pmatrix} \ .
\end{eqnarray}
Also, making use of  the continuity condition at $x^{iv}=0$, i.e., at  $x=(2d+w)$,
we obtain

\begin{equation}
{\cal N}_1\begin{pmatrix}G^{iv}\cr
H^{iv}\end{pmatrix}={\cal M}_1\begin{pmatrix}J^{iv}\cr
K^{iv}\end{pmatrix}
\end{equation}
and thus

\begin{equation}
\begin{pmatrix}G\cr H\end{pmatrix}={\cal T}_{3V}^{-1} .{\cal N}_1^{-1}. {\cal M}_1.{\cal T}_{30}.\begin{pmatrix}J\cr
K\end{pmatrix} \ ,
\end{equation}
where

\begin{eqnarray}
{\cal T}_{30}&=&\begin{pmatrix} e^{iq_{1,x}(2d+w)}&&0 \\
0&&e^{-iq_{1,x}(2d+w)} \end{pmatrix}
\nonumber\\
{\cal T}_{3V}&=&\begin{pmatrix} e^{iq_{2,x}(2d+w)}&&0 \\
0&&e^{-iq_{2,x}(2d+w)} \end{pmatrix}\ .
\end{eqnarray}
So,  we have a relationship between  the pairs of coefficients $(A,B)$ and $(J,K)$ given by

\begin{equation}
\begin{pmatrix}A\cr B\end{pmatrix}=({\cal M}_1^{-1}  {\cal N}_1)\cdot  ({\cal T}_{10}^{-1} 
{\cal M}_1^{-1}  {\cal N}_1  {\cal T}_{1V})^{-1} \cdot ({\cal T}_{20}^{-1}  {\cal M}_1^{-1}
{\cal N}_1  {\cal T}_{2V})\cdot
({\cal T}_{30}^{-1}  {\cal M}_1^{-1}  {\cal N}_1  {\cal T}_{3V})^{-1}\cdot
\begin{pmatrix}J  \cr  K\end{pmatrix} \  
\end{equation}
By induction,      we have for $(N+1)$   potential barriers

\begin{eqnarray}
&&
\begin{pmatrix}1\cr
B\end{pmatrix}
=({\cal T}_{00}^{-1}  {\cal M}_1^{-1}  {\cal N}_1  {\cal T}_{0V})\cdot ({\cal T}_{10}^{-1}
 {\cal M}_1^{-1}  {\cal N}_1  {\cal T}_{1V})^{-1}\cdot
({\cal T}_{20}^{-1}  {\cal M}_1^{-1}  {\cal N}_1  {\cal T}_{2V})
\nonumber\\
&\times&({\cal T}_{30}^{-1}  {\cal M}_1^{-1}  {\cal N}_1  {\cal T}_{3V})^{-1}
\cdots({\cal T}_{(2N+1)0}^{-1}  {\cal M}_1^{-1}  {\cal N}_1  {\cal T}_{(2N+1)V}^{-1})
\begin{pmatrix}t\cr
0\end{pmatrix}
\end{eqnarray}
Generalizing our formalism to $N+1$ equally spaced potential barriers,
  our calculations show that

\begin{equation}
{\cal T}_{(2N+1)0}=\begin{pmatrix} e^{iq_{1,x}\left\{(N+1)d+Nw\right\}}&&0 \\
0&&e^{-iq_{1,x}\left\{(N+1)d+Nw\right\}} \end{pmatrix}
\end{equation}

\begin{equation}
{\cal T}_{(2N+1)V}=\begin{pmatrix} e^{iq_{2,x} \left\{(N+1)d+Nw\right\} }&&0 \\
0&&e^{-iq_{2,x} \left\{(N+1)d+Nw\right\}} \end{pmatrix} \ .
\end{equation}
 Therefore, solving this equation for  $t$,  we obtain the transmission
coefficient for multibarrier \cite{multi1,multi2} as $T=t\cdot t^\ast$. We now use these results to carry out our
numerical calculations.

\subsection{Conductance}

The conductance  coefficient\cite{cond1} for a specific spin direction and valley across a
potential step or potential barrier  is given in terms of the transmission probability
as follows:

\begin{equation}
G=\frac{e^2}{h}\int_0^{2\pi} d\theta\ T(\theta) \cos(\theta) \ .
\end{equation}
We shall employ this result below to investigate the behavior of the conducting
properties of graphene under the influence of a split gate to produce an $n-p$ or $n-p-n$
junction. However, we must first analyze the transmission coefficient which we derived
above in our formalism for gapped graphene.

\medskip
\par

\section{Numerical Results and Discussion}
\label{sec3}

\begin{figure}[h]
\centering
\includegraphics[width=2.87in,height=3.21in,keepaspectratio]{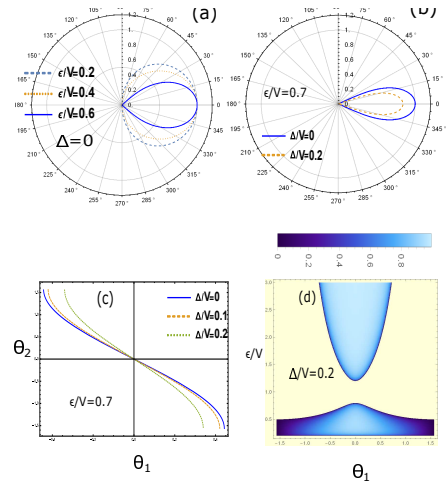}
\caption{(Color online) Polar plots of the transmission coefficient at
different incident energies for (a) gapless and (b) gapped graphene for
chosen incident energy.  In (c), we show the relationship between the angle
of incidence and the angle of refraction for chosen energy gap and incident
energy. Density plot of the transmission coefficient across a step is shown in (d)
as a function of the angle of incidence and energy. }
\label{FIG:1}
\end{figure}

 In Figs.\ \ref{FIG:1}(a) and \ref{FIG:1}(b), we present polar  plots  for  the
transmission probability across a potential  step for chosen incident energy in units of the
barrier height and compare results for gapless and gapped graphene. Klein tunneling
is evident at normal incidence when $\Delta=0$ but there is some reflection in  the
case for gapped graphene. Off normal incidence, the transmission probability for gapless and
gapped graphene again shows some differences which may also be significant,
compared to head-on incidence for the electron beam.  In Fig.\ \ref{FIG:1}(c), we  also plot the
 angle of incidence as a function of the corresponding angle of refraction
  for chosen energy gap, thereby demonstrating the way in which negative refraction
 is affected by an energy gap in monolayer graphene. We also exhibit in Fig.\ \ref{FIG:1}(d) a density
plot for the transmission probability across a potential step as a function of incident
angle and energy for   gapped   monolayer graphene.
Our results clearly show that there is a forbidden energy region  for incoming electrons
  at any angle and that these results differ substantially from those corresponding
 to tunneling across a potential barrier\cite{Negrf8}.

\medskip
\par

\begin{figure}[h]
\centering
\includegraphics[width=2.87in,height=3.21in,keepaspectratio]{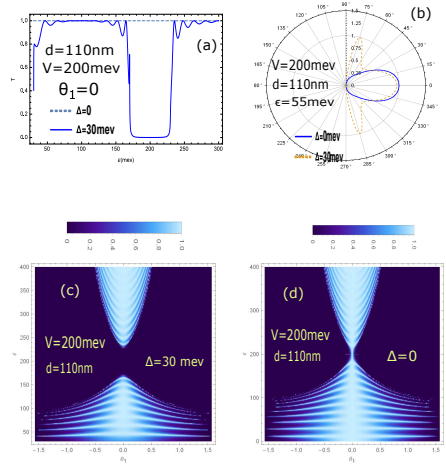}
\caption{(Color online) (a) Plots for the transmission probability  of   an electron through
a single potential barrier versus the incident energy. In (b), polar plots  are
presented for transmission for different angles of incidence. Panels (c) and
(d) respectively show   transmission density plots for   gapped and gapless graphene.}
\label{FIG:2}
\end{figure}

Figures \ref{FIG:2}(a), \ref{FIG:2}(b), \ref{FIG:2}(c) and \ref{FIG:2}(d)   illustrate
 the changes in the transmission probability
 for different  incident energy of an electron impinging on a single potential barrier for
 gapped and gapless graphene with chosen barrier width. Unlike gapless graphene, for
normal incidence in gapped graphene,   the transmission probability  oscillates
as the energy is varied in Fig.\
\ref{FIG:2}(a) and its value drops drastically when the incident energy is
 close to the barrier height. The polar plots in the Fig.\ \ref{FIG:2}(b) and
 for the chosen barrier width show  that   Klein tunneling is suppressed at normal incidence, thereby
illustrating   some possible reflection even for zero angle of incidence.  We also present
in Figs.\ \ref{FIG:2}(c), (d)  the transmission results in the form
of density plot as a function of incident energy and the angle of incidence. The results
show the brighter region corresponding to  higher transmission probability and there is
the region in the middle of the figure where transmission is not allowed for some values
of energy for any angle of incidence when there is a band gap.

\medskip
\par

\begin{figure}[H]
\centering
\includegraphics[width=2.87in,height=3.21in,keepaspectratio]{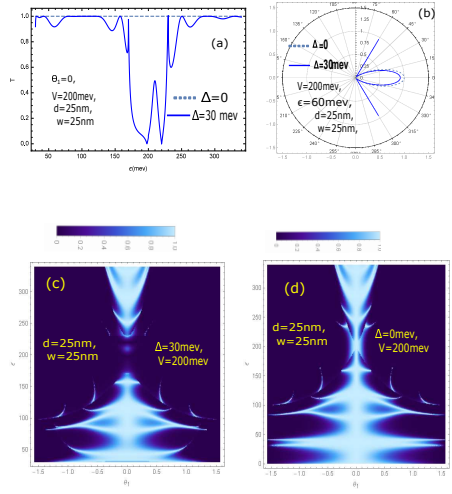}
\caption{(Color online)    Transmission probability through
two potential barriers. All   parameter for the barrier   height,
and energy band gap   are the same as in Fig.\  \ref{FIG:2}.
But, here  the barrier width  $d$=$50nm$ and the inter-barrier separation $w$=$50nm$.}
\label{FIG:4}
\end{figure}

We now turn our attention to the transmission of an electron through two electrostatic
potential barriers separated by some distance $w$. Our results are presented in Figs.\
 \ref{FIG:4}(a), \ref{FIG:4}(b), \ref{FIG:4}(c) and \ref{FIG:4}(d). The inter-barrier
 separation leads to  polar
plots which, as expected,  bear similarities with that obtained in Fig.  \ref{FIG:2},
with forbidden regions of transmission for some range of energy and incident angle.
 However, unlike the case for a single barrier, we observe
some bright    regions in the middle section of the density plot. These spikes
are due to the commensurability oscillations  of the waves in between the barriers.
The spike locations depend  on the   parameters chosen for the barrier
width as well as the energy band gap.

\medskip
\par

We have made  use of our results for the transmission probability to calculate the
ballistic conductance  for all the above mentioned geometries. For a  potential step,
 Fig.\  \ref{FIG:6}(a) shows the conductance as a function of the energy gap parameter
 $\Delta$ for    chosen values of incoming electron energy. Here, the conductance is almost
 constant for  small $\Delta/V$  and it falls  precipitously when
 the band gap parameter is close to the electron incident energy.
 Figure \ref{FIG:6}(b) shows the variation of the conductance coefficient with incident
 electron  energy for various values of band gap.  For the range of energy
 shown, the conductance decreases monotonically with energy when $\Delta=0$.
 However, when there is an energy band gap, there is conductance only when $\epsilon > \Delta$
 as required for graphene electrons in the conduction band. This requirement
 leads to the expected behavior of conductance in Figs. \ref{FIG:6} (a) and (b).

\medskip
\par

We vividly demonstrate in Fig.\ \ref{FIG:7}  that there is a relationship
between the electron energy  $\epsilon$ and band gap parameter $\Delta$
showing the region in $\epsilon-\Delta$ space when there is focusing
 of an electron beam across a potential step in monolayer graphene.  Thus, we have identified
a limited region where the refractive index   is negative.  Outside this region,
the refractive index may either be positive (transmission without focusing) or
there is no transmission at all, i.e.,     the equation governing refraction
may not be satisfied, to yield real solutions for Eq.\ (\ref{angle1}). In particular, these results
show that for gapless graphene,  the range of energy for negative refraction
is larger than for gapped graphene. This dramatic result from our theory
means that one may specifically choose   negative refractive index by allocating the band gap.

\medskip
\par

\begin{figure}[H]
\centering
 \includegraphics[width=2.87in,height=3.1in,keepaspectratio]{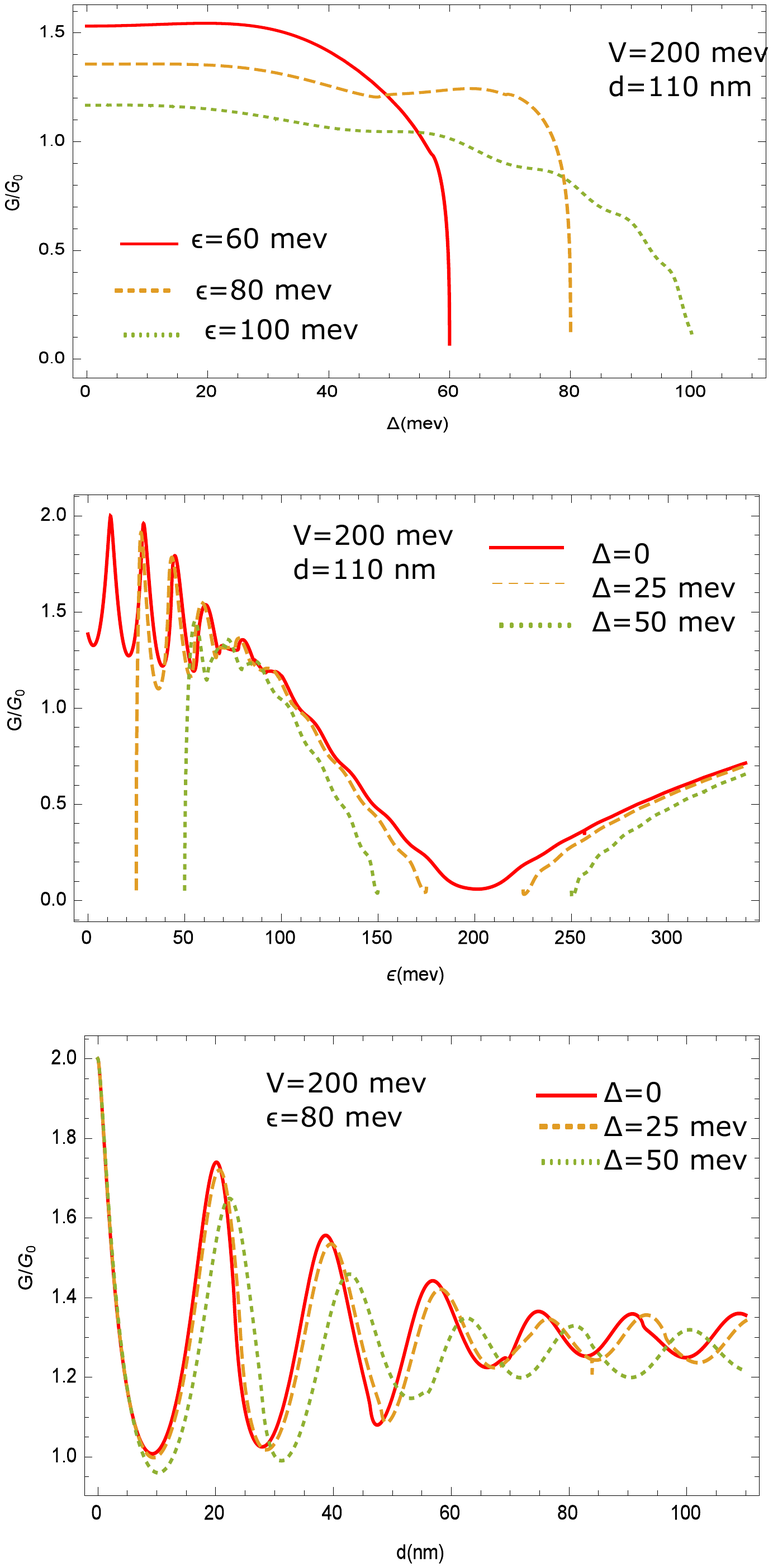}
\caption{(Color online) Conductance plots for ballistic transport
for gapped and gapless graphene when there is only one potential barrier.
The conductance coefficient is plotted as a function of  (a) the
energy band parameter $\Delta$, (b)  incident energy $\epsilon$, and (c)
potential barrier width $d$. }
\label{FIG:8}
\end{figure}

We present novel results for the $n-p-n$ conductance as functions  of  the
energy gap parameter $\Delta$, the incident energy $\epsilon$  and the barrier
width $d$ in Figs.\ \ref{FIG:8}(a), \ref{FIG:8}(b) and \ref{FIG:8}(c), respectively.  Unlike the
results in Fig.\ \ref{FIG:6} for a potential step, in an $n-p-n$ junction,
the conductance in Fig.\ \ref{FIG:8}(b) has characteristic commensurability
oscillations due to the ratio between the de Broglie wavelength and the barrier
width.  The reflections at the barrier interface  change a smooth curve in Fig.\ \ref{FIG:6}(b)
 into a decreasing, oscillating curve in Fig.\ \ref{FIG:8}(b)  for the conductance
 as a function of the energy parameter $\epsilon$.   It is evident
 that commensurability oscillations for an $n-p-n$ junction are superimposed  on
 the $n-p$ junction conductance results, including a lower bound on the existence of this
 transport coefficient. We also note that  Fig.\ \ref{FIG:8}(b)  shows  that for
 finite energy gap, there is a range of energy for which the conductance is zero
 and that this range of energy for an insulating phase increases with $\Delta$.
 Figure\ \ref{FIG:8}(c)   clearly demonstrates the commensurability oscillations as the
 barrier width is varied.

 \medskip
 \par

\begin{figure}[H]
\centering
\includegraphics[width=4.67in,height=2.41in,keepaspectratio]{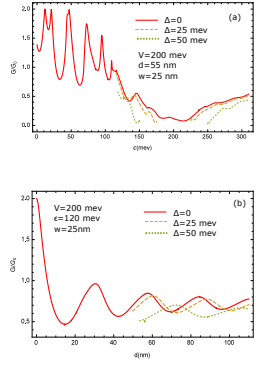}
\caption{(Color online) Conductance plots,  in units of $G_0=e^2/h$,
 for gapped and gapless graphene when there is a pair of potential barriers.
Panel  (a) shows the conductance as a function of the  incident electron energy
for chosen band gap with barrier width  $d$=$55 nm$ and inter-barrier separation
$w$=$25$ nm. Panel (b)shows conductance versus barrier width for chosen parameter values
 as indicated in the figure.}
\label{FIG:9}
\end{figure}

Based on the formula for the conductance, we carried out calculations  across two potential
 barriers  as shown in Figs.\ \ref{FIG:9}(a) and \ref{FIG:9}(b). As the energy is increased,
 the  conductance shows a periodic oscillations and we   observe a dip in the conductance
 when the electron energy is close  to the barrier height. This is because of the drop
in the transmission probability in that regime.  As for an $n-p-n$ junction,
the conductance does not exist for energy less than the band gap parameter $\Delta$
and for the difference between the electron  energy and the barrier height less
than   $ \Delta$.  Figure\ \ref{FIG:9}(b) shows the   conductance  versus barrier width. As
 in the case of an  $n-p-n$ junction, the conductance  falls off
 from a high  to a reduced value for the lower barrier width. As the width is increased,
 the conductance  oscillates as shown in Fig.\   \ref{FIG:9}(b). When the band gap is finite,
 the average conductance is decreased   as the band gap is increased.

\medskip
\par

\section{Concluding Remarks}
\label{sec4}

The interaction between Dirac particles in graphene and circularly polarized light
 leads to the formation of quantum electron dressed states with an energy band
gap of order 50 meV. This gap may   be produced by the application of laser light of
suitable power. So, because of the creation of a  band gap  between the valence and
conduction bands  in graphene, we do not see the Klein tunneling for  normal incidence
of an electron beam. Instead, we observe complete transmission at some other angle of
 incidence. When there are  two or more potential barriers present, we observe that
the number of peaks in the transmission spectrum increases. This is due to interference
 between oppositely traveling waves  within a barrier and between separated barriers.
The interference effect  is reduced when the barriers are brought closer together.

\medskip
\par
\begin{figure}[h]
\centering
\includegraphics[width=5in, keepaspectratio]{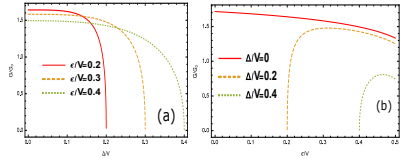}
\caption{(Color online) Conductance plots,  in units of $G_0=e^2/h$,
 for gapped and gapless graphene when there is a step potential.  Panel (a)
 shows the conductance versus the band gap parameter for  chosen energy.
 Panel (b) presents  how  the conductance varies with incident energy for chosen band gap.  }
\label{FIG:6}
\end{figure}
\begin{figure}[h]
\centering
\includegraphics[width=4.67in,height=2.41in,keepaspectratio]{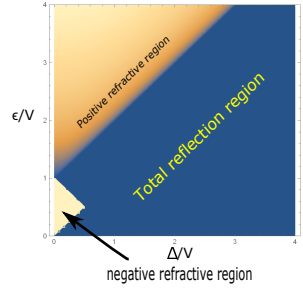}
\caption{(Color online) Density plot of limited   negative refractive index  as
a function of incident energy and energy band gap. }
\label{FIG:7}
\end{figure}

An important result coming out of our calculations is that there is a well defined
region in $\epsilon-\Delta$  space where negative refraction, i.e., focusing
of an electron beam,  may occur for monolayer    graphene when a negative-positive
split gate is applied . Outside this region, the beam may not be refracted at all or,
if there is refraction, the refractive index may be positive.   This means that
by varying the energy band gap appropriately, we could have a selected
refractive index for monolayer graphene.  Our conductance calculations show that there is
a lower bound for the incident energy for gapped graphene  with a $p-n$ junction, thereby
indicating a metal-insulator transition. The regions of finite refraction in Fig.\
\ref{FIG:7} determine the range for the incident energy and gap for
monolayer graphene   where the conductance is finite  in Fig.\ \ref{FIG:6}(a).
When a potential barrier is introduced, our calculations show that  there are
commensurability oscillations superimposed on the conductance curves for a step
due to interference between incident and reflected waves from the potential walls.
Therefore, our present work plays  an important  role in demonstrating that monolayer graphene  
may serve as a metamaterial only in a well-defined range of electron energy and band gap,
i.e., focusing may be achieved in a  limited region of $\epsilon-\Delta$ space
when a positive-negative bias is applied.  The conductance for monolayer graphene with
an electrostatic barrier or multiple barriers reveals interference  effects  due to the 
ratio of the de Broglie wavelength to the barrier width.

\medskip
\par

\newpage

\end{document}